\documentstyle[12pt,epsf]{article}

\newcommand{\be}{\begin{equation}}
\newcommand{\ee}{\end{equation}}
\newcommand{\bea}{\begin{eqnarray}}
\newcommand{\eea}{\end{eqnarray}}
\newcommand{\ra}{\rightarrow}
\newcommand{\lan}{\langle}
\newcommand{\ran}{\rangle}
\begin{document}

\font\fortssbx=cmssbx10 scaled \magstep2
\hbox to \hsize{
\includegraphics{uwlogo.ps}
\hskip.5in \raise.1in\hbox{\fortssbx University of Wisconsin - Madison}
\hfill$\vcenter{\hbox{\bf MAD/PH/851}
            \hbox{October 1994}}$ }
\vskip 2cm
\begin{center}
\Large
{\bf Relativistic Flux Tube Model Calculation of the Isgur-Wise Function} \\
\vskip 0.5cm
\large
M. G. Olsson and Sini\v{s}a Veseli \\
\vskip 0.1cm
{\small \em Department of Physics, University of Wisconsin, Madison,
	\rm WI 53706}
\end{center}
\thispagestyle{empty}
\vskip 0.7cm
\begin{abstract}
The Relativistic Flux Tube model is used to  calculate
of the Isgur-Wise functions describing the exclusive
semileptonic
decays of  $\bar{B}$ and $\bar{B}_{s}$ mesons. The light quark mass
dependence is investigated and the predicted universal
function agrees well with  the results of lattice
simulations, and with the experimental data.
Recent experimental measurements of the
$\bar{B}\ra D^{*}l\bar{\nu}_{l}$
 decay
distribution yield the CKM element $V_{bc}=0.035\pm 0.001$.
The IW function slope and second derivative
at the zero recoil point are predicted to be
$\xi'(1) = -0.93\pm 0.04$ and
$\xi''(1)= 1.7\pm 0.1$ for a range of light quark massses.
The importance of including higher derivatives
in analyses of experiment is emphasized.
\end{abstract}

\newpage
\vspace*{+1mm}
\section{Introduction}
\vspace*{+3mm}
In the
heavy quark limit (mass of the heavy quark
$m_{Q}\hspace{-2pt}\ra\hspace{-2pt}\infty$)
all the non-perturbative,
strong interaction physics for semileptonic $\bar{B}\ra Dl\bar{\nu}_l$
and $\bar{B}\ra D^{*}l\bar{\nu}_{l}$ decays (and corresponding
$\bar{B}_{s}$ decays), can be parametrized in terms of a
single universal function \cite{bib:isg-wis}, known as the Isgur-Wise (IW)
function. Knowledge of the IW function is essential in all
calculations that lead to numerical values of
branching ratios of semileptonic decays and of the
$V_{cb}$ element of the CKM matrix. A reliable
calculation of the IW function from  first principles is
still beyond our power, although recent efforts
using lattice QCD are very encouraging. A variety of different
meson  models
make specific predictions for the IW function. Since these model results vary
widely, the IW prediction can be considered to be a sensitive
gauge of  model validity.

The relativistic flux tube (RFT) model shows promise to
provide a realistic description of all meson states. The RFT
model is in essence a description of dynamical confinement
\cite{bib:dan}-\cite{bib:sik}. For slowly moving quarks rigorous QCD
relativistic corrections \cite{bib:ei-fe}-\cite{bib:br-pr}
clearly demonstrate
that the scalar confinement potential picture is incorrect
\cite{bib:col-ken,bib:br-pr}. On the other hand, the RFT
model dynamics is
consistent with both spin-dependent \cite{bib:ken,bib:buch,bib:pi-st}
and spin-independent \cite{bib:dan,bib:col-ken} QCD expectations.
In this paper we examine the RFT model predictions for IW functions.
We find here that the spinless light degrees of freedom model
is in excellent agreement with both experiment and with fundamental QCD
expectations.

We begin by a brief review of the RFT formalism. In Section
3 we outline the general calculation of the IW function and its
derivatives at the zero recoil point. Our numerical
results and conclusions are given
in Section 4. Here we evaluate the RFT predictions for the IW function
and examine its dependence on the light quark mass. We  determine
the $V_{cb}$ CKM element and branching fractions for various
exclusive semi-leptonic decays and  test
the shape of the predicted IW function. Finally we note that the RFT
prediction for the slope at the zero recoil point is in good
agreement with present lattice simulation results.

\section{RFT model in the heavy quark limit}

In the limit
$m_{\bar{Q}}\ra\infty$ the quantized equations of the spinless RFT model
with light quark mass $m_{q}$ are \cite{bib:col-dan,bib:sik}
\bea
\frac{\sqrt{l(l+1)}}{r}&=&\frac{1}{2}\{W_{r},\gamma_{\perp}v_{\perp} \}
+ a \{r,f(v_{\perp})\}\ ,\label{eq:rftj}\\
H_{q}=H-m_{\bar{Q}}&=&\frac{1}{2}\{W_{r},\gamma_{\perp}\}+
\frac{a}{2}\{r,\frac{\arcsin{v_{\perp}}}{v_{\perp}}\}
+V(r)\ . \label{eq:rfth}
\eea
Here $H_{q}$ is Hamiltonian of the light degrees of freedom
(LDF),
with the short range potential
\be
V(r) = -\frac{\kappa}{r}\ ,
\ee
 and
\bea
 W_{r}&=&\sqrt{p_{r}^{2}+m_{q}^2}\ , \\
\gamma_{\perp}&=&\frac{1}{\sqrt{1-v_{\perp}^2}}\ , \\
 f(v_{\perp})&=&\frac{1}{4 v_{\perp}}
(\frac{\arcsin{v_{\perp}}}{v_{\perp}}-\frac{1}{\gamma_{\perp}})\ .
\eea
It is assumed that $v_{\perp}$ is Hermitian, and the symetrization
(curly brackets in  (\ref{eq:rftj}) and (\ref{eq:rfth}))
then yields a Hermitian Hamiltonian.

The solution of the energy eigenvalue
equation
\be
H_{q}\Phi = E_{q}\Phi \label{eq:heq}
\ee
required the development of methods to eliminate
the a-priori unknown operator $v_{\perp}$ and also to address
the solution of  operator equations in which functions
of both $p_{r}$ and $r$ appear. Our approach
\cite{bib:dan,bib:col-dan,bib:sik} employs the Galerkin method
\cite{bib:galerkin}
in which a truncated basis set is used to transform the operator equations
(\ref{eq:rftj}-\ref{eq:heq}) into finite matrix equations.
The resulting matrix equations describe the motion of a
light quark with radial momentum $p_{r}$ and perpendicular
velocity $v_{\perp}$ relative to a
static heavy antiquark. In order to solve
for the eigenvalues of the Hamiltonian, one first solves
the angular momentum equation (\ref{eq:rftj}) for the $v_{\perp}$ matrix,
and then diagonalizes
the  $H_{q}$ matrix (\ref{eq:rfth}) as discussed in
detail in references \cite{bib:col-dan,bib:sik}.

Since there is no spin-orbit coupling in the above model, we can
write our wave function (describing the LDF
 in the meson rest frame) as a product of the orbital wave function
and the light quark spinor as
\be
\Psi^{(0)}_{H_{\bar{Q}}}(x) = \Phi^{(0)}_{\alpha LM_{L}}({\bf x})
\phi^{(0)}_{m_{s_{q}}}e^{-iE_{q}t}\ .
\ee
Here, $H_{\bar{Q}}$ represents  all quantum numbers of
the LDF, and the superscript $0$
denotes the rest frame of a meson.

\section{Decays $H_{\bar{b}}\ra H_{\bar{c}}l\bar{\nu}_l$ and the
IW functions in the spinless RFT model}

For heavy-light mesons,
the spin of the heavy quark ${\bf s}_{\bar{Q}}$ and the spin of the light
antiquark ${\bf s}_{q}$ decouple as $\frac{1}{m_{\bar{Q}}}$, and
in the limit $m_{\bar{Q}}\ra\infty$ they are conserved separately
by the strong interactions. Because
of this, hadrons containing a single heavy quark can be
simultaneously assigned the quantum numbers
$(s_{\bar{Q}},m_{s_{\bar{Q}}})$ and $(s_{q},m_{s_{q}})$, and
matrix element
of the hadron current describing semileptonic decay can
be expressed by the free heavy quark current and the IW function.

Except for a trivial kinematical factor, the IW function
$\xi(\omega)$ is defined \cite{bib:li-ho,bib:sa-za}
as the
overlap of the wave functions describing the light degrees of freedom
in the two mesons, i.e.
\be
\xi(\omega) = \sqrt{\frac{2}{\omega + 1}}\
\langle \Psi_{H_{\bar{c}}}|\Psi_{H_{\bar{b}}}\rangle \ .
\label{eq:iwf}
\ee
Here, if the meson velocity  in the lab frame is ${\bf v}$,
the wave function
describing its LDF is
\be
\Psi_{H_{\bar{Q}}}(x')=S({\bf v})\Psi_{H_{\bar{Q}}}^{(0)}(x)\ ,
\label{eq:boost}
\ee
with $x'=\Lambda^{-1}({\bf v})x$ being the lab frame, $x$
the rest frame of the particle, and $S({\bf v})$ is the
wave function
Lorentz boost.

Since the IW function is Lorentz invariant, we can choose any frame
to calculate it. Particularly convinient is
the modified Breit frame
\cite{bib:li-ho,bib:sa-za}, where the two particles move
along the z-axis with equal and
opposite velocities. Denoting  velocity of $H_{\bar{c}}$ as ${\bf v}$,
by the use of (\ref{eq:boost}) the
overlap takes the form
\bea
\langle \Psi_{H_{\bar{c}}}|\Psi_{H_{\bar{b}}} \rangle &=&
\langle \Phi_{H_{\bar{c}}}|\Phi_{H_{\bar{b}}}\rangle\
\langle \phi_{H_{\bar{c}}}|\phi_{H_{\bar{b}}}\rangle \nonumber \\
\vspace*{+2mm}
&=&
\langle \phi_{H_{\bar{c}}}|\phi_{H_{\bar{b}}}\rangle
\int d^{3}x' \Phi_{H_{\bar{c}}}^{\dagger}(x')\Phi_{H_{\bar{b}}}(x')|_{t'=0} \\
\vspace*{+2mm}
&=&
\langle \phi_{H_{\bar{c}}}^{(0)}|S^{\dagger}({\bf v})
S(-{\bf v})|\phi_{H_{\bar{b}}}^{(0)}\rangle
\int d^{3}x' \Phi_{H_{\bar{c}}}^{(0)\dagger}(x_{+})
S^{\dagger}({\bf v})S(-{\bf v})\Phi_{H_{\bar{b}}}^{(0)}(x_{-})|_{t'=0}\ .
\nonumber
\eea
 In this expression
$x_{+}$ and $x_{-}$ denote the rest frames
of $H_{\bar{c}}$ (moving in the +z direction) and $H_{\bar{b}}$
(moving in the -z direction), respectively. Using the fact
that Lorentz boosts are real, i.e.
$S^{\dagger}({\bf v})=S({\bf v})=S^{-1}(-{\bf v})$, the
boost factors
cancel out. The overlap of spin part of the wave functions will just give us
$\delta_{m_{s_{q}}^{H_{\bar{c}}},m_{s_{q}}^{H_{\bar{b}}}}$,
which we suppress in the following, and
we are left with
\be
\langle \Psi_{H_{\bar{c}}}|\Psi_{H_{\bar{b}}} \rangle =
\int d^{3}x' \Phi_{H_{\bar{c}}}^{(0)\dagger}(x_{+})
\Phi_{H_{\bar{b}}}^{(0)}(x_{-})|_{t'=0} \ .
\ee
Denoting $v=|{\bf v}|$ and $\gamma=\frac{1}{\sqrt{1-v^{2}}}$, and using
\be
x_{\pm}|_{t'=0}=\Lambda(\pm{\bf v})x'|_{t'=0}
=(\mp \gamma vz',x',y',\gamma z') \ ,
\ee
we  obtain
\bea
\langle \Psi_{H_{\bar{c}}}|\Psi_{H_{\bar{b}}} \rangle &=&
\int d^{3}x' \Phi_{H_{\bar{c}}}^{(0)\dagger}(-\gamma vz',x',y',\gamma z')
\Phi_{H_{\bar{b}}}^{(0)}(+\gamma vz',x',y',\gamma z') \nonumber \\
\vspace*{+2mm}
&=&
\int d^{3}x' \Phi_{H_{\bar{c}}}^{(0)\dagger}(x',y',\gamma z')
\Phi_{H_{\bar{b}}}^{(0)}(x',y',\gamma z')
e^{-2iE_{q}\gamma vz'} \ ,
\eea
since the energies of the light degrees of freedom are the same in both
hadrons. Finally, after rescaling the $z'$ coordinate
($z'\ra\frac{1}{\gamma}z'$), renaming integration variables, and noting
kinematical identities
\bea
\gamma=\sqrt{\frac{w+1}{2}}\ , \\
v=\sqrt{\frac{\omega-1}{\omega+1}} \ , \label{eq:kinv}
\eea
valid in the Breit frame, the IW function becomes \cite{bib:sa-za}
\be
\xi(\omega) =
\frac{2}{\omega + 1}
\int d^{3}x \Phi_{H_{\bar{c}}}^{(0)\dagger}({\bf x})
\Phi_{H_{\bar{b}}}^{(0)}({\bf x})
e^{-2iE_{q} vz} \ .
\ee

By spherical symmetry, and since we are interested here
only in  s-waves, the orbital wave function has the form
\be
\Phi_{\alpha 00}({\bf x})=R_{\alpha 0}(r)\frac{1}{\sqrt{4\pi}}
\ .
\ee
 For the calculation of the IW function we
need to evaluate integral of the form
\be
\frac{1}{4\pi}\int d^{3}x
R_{\alpha' 0}^{*}(r)R_{\alpha 0}(r)
 e^{-2iE_{q} vz}\ .
\ee
Using
\be
e^{-ikz}=\sum_{l=0}^{\infty}(2l+1)(-i)^{l}j_{l}(kr)
\sqrt{\frac{4\pi}{2 l+1}}Y_{l0}\ ,
\ee
and
orthonormality of the spherical harmonics,
one easily obtains
 the IW function in the form
\be
\xi(\omega)
=
\frac{2}{\omega + 1}
\lan\
j_{0}(2E_{q}\sqrt{\frac{\omega-1}{\omega+1}}r)
\ran\ , \label{eq:iwfgr}
\ee
where
\be
\lan A \ran =
\int_{0}^{\infty}dr\  r^2 R_{\alpha'0}(r)A(r)R_{\alpha 0}\ .
\ee

The result (\ref{eq:iwfgr}) can also be used
to obtain derivatives
of the IW function at the zero recoil point ($\omega = 1$).
Explicit results for the first two such derivatives are
\bea
\xi'(1)& =& -(\frac{1}{2}+\frac{1}{3}E_{q}^{2}
\lan r^{2}\ran) \ ,\label{eq:first}\\
\xi''(1)& =& \frac{1}{2}+\frac{2}{3}E_{q}^{2}\lan r^{2}\ran
+\frac{1}{15}E_{q}^{4}\lan r^{4}\ran\ .\label{eq:second}
\eea
Since all integrals in these expressions are positive, we recover
the limit \cite{bib:sa-za}
\be
\xi'(1)<-\frac{1}{2}\ ,
\ee
and also obtain
\be
\xi''(1)>\frac{1}{2}
\ee
Similar limits can be found for the third and higher order
derivatives.

\section{Results}

The parameters appearing in the spinless RFT model are
the string tension $a$,   the short range potential constant $\kappa$,
and the quark masses. In
our calculation, we fix the tension from the universal Regge slope
$\alpha ' \simeq =0.8\ GeV^{-2}$,
\be
a=\frac{1}{2\pi \alpha'}\simeq 0.2 GeV^{2}\ .
\ee
This is consistent with the value found from analyses of heavy onia
spectroscopies \cite{bib:galerkin}. We also fix the light quark mass
$m_{u,d}$, as the quality of
our fit depends only weakly on its value. The remaining parameters
$\kappa$, $m_{s}$, $m_{c}$ and $m_{b}$ are varied
to best account
for the  6 spin averaged
states \cite{bib:col-dan,bib:sik}. As an example, in Table \ref{tab:hl} we
show the results of our fit when $m_{u,d}$ was fixed at $0.3\  GeV$.
Once the parameters of the model are determined,
the wave functions and LDF energies $E_{q}$ are known and the
IW function can be calculated with the aid
of  (\ref{eq:iwfgr}).

The IW function, as computed from  (\ref{eq:iwfgr}), is displayed
in Fig. \ref{fig:iwfb}. The solid and dashed curves
correspond to $m_{u,d}=0$ and $300\ MeV$ respectively.
We observe that with increasing light quark mass
the two IW functions differ by only  a few percent at
 $\omega = 1.5$.

One can expand the IW function
about zero recoil point as discussed
at the end of the preceeding section. Using (\ref{eq:first})
and (\ref{eq:second}) we have
explicitly
evaluated the first and second
derivatives over the usual range
of light
quark masses $m_{u,d}$
as shown in Figs. \ref{fig:sl} and \ref{fig:sd}. These derivatives
vary by less than
ten percent over the range of $m_{u,d}$
and we conclude that
\bea
\xi'(1)&=&-0.93\pm 0.04\ , \label{eq:numsl} \\
\xi''(1)&=&1.7\pm 0.1\ .
\eea
In Fig. \ref{fig:ps} we compare the
 power series approximation
about the zero recoil point
$\omega = 1$
to the full IW function. We observe
that the power series expansion
is not very
convergent and that a significant error is
encountered even including cubic terms
in $\omega-1$ near the high end of the $\omega$ range. In particular,
if one truncates
the series to include only the linear
term, then in fitting to the data the slope will be overestimated, i. e.
the true slope will be more negative.

The explicit values of the derivarives will provide
an excellent
test of the RFT model through lattice simulation
calculations. It has been proposed \cite{bib:aglietti} that a direct
evaluation
of higher derivatives
of the IW function
at zero recoil point
is possible by numerical QCD methods.

In Fig. \ref{fig:iwfbs}
we show the ratio of the IW functions for $\bar{B}_{s}$
and $\bar{B}$ semileptonic decays. Using light quark mass
 values $m_{s}=513 \ MeV$ and $m_{u,d}=300 \ MeV$,
from the fit of Table \ref{tab:hl}, we see
that the IW function for the two decays
differ by less than by 1.5\%. We also note that
$\xi_{\bar{B}_{s}}(\omega)\leq \xi_{\bar{B}}(\omega)$.

A comparison of our IW function $\xi_{\bar{B}}(\omega)$ with
the recent ARGUS \cite{bib:arg}
 and CLEO II data \cite{bib:cleo} (shown
in Fig. \ref{fig:arg}) yields the $V_{cb}$ element
of the CKM matrix
\be
V_{cb}=0.035\pm 0.001
\ee
(the $\chi^2$ of this
fit was 0.88 per degree of freedom).
The RFT prediction for the
$\omega$ dependence of the IW function is in excellent agreement
with the data.
Finally, using expressions given in \cite{bib:neudif} for differential
widths of $\bar{B}$ and $\bar{B}_{s}$ decays, we  calculate
corresponding branching ratios to be
\bea
Br(\bar{B}\ra Dl\bar{\nu}) &=&
1.79(\frac{\tau_{\bar{B}}}{1.53ps})\% \ ,\\
Br(\bar{B}\ra D^{*}l\bar{\nu}) &=&
5.01(\frac{\tau_{\bar{B}}}{1.53ps})\% \ , \\
Br(\bar{B}_{s}\ra D_{s}l\bar{\nu}) &=&
1.83(\frac{\tau_{\bar{B}_{s}}}{1.53ps})\% \ , \\
Br(\bar{B}_{s}\ra D^{*}_{s}l\bar{\nu}) &=&
5.11(\frac{\tau_{\bar{B}_{s}}}{1.53ps})\% \ .
\eea
Agreement with the available data \cite{bib:pdg}
for $\bar{B}\ra D$  $(1.8\pm0.4)\%$
and for $\bar{B}\ra D^{*}$ $(4.5\pm0.4)\%$ is
satisfactory. We also find that
\be
Br(\bar{B}\ra Dl\bar{\nu}) \approx Br(\bar{B}_{s}\ra D_{s}l\bar{\nu})\ ,
\ee
and
\be
Br(\bar{B}\ra D^{*}l\bar{\nu}) \approx
Br(\bar{B}_{s}\ra D^{*}_{s}l\bar{\nu})\ .
\ee
as expected from the heavy-quark symmetry.

Since the IW function is normalized at the zero recoil point to
$\xi(1)=1$, much of the model dependence is distilled into the
slope at zero recoil $\xi'(1)$. Various models have
yielded a wide range of slopes \cite{bib:slopes} from
$-\frac{1}{3} > \xi'(1) > -2$. Our result for the
slope  in the RFT model is from (\ref{eq:numsl})
\be
\xi^{'}_{\bar{B}}(1)=-0.93 \pm 0.04\ ,
\ee
where the error is estimated from the  slope dependence
on the light quark mass shown in Fig. \ref{fig:sl}.
The recent CLEO II experimental slope value \cite{bib:cleo} is
\be
\xi^{'}_{\bar{B}}(1)=-0.82 \pm 0.12 \pm 0.12\ ,
\ee
which is consistent with our predicted value.
This value was obtained by assuming
$\xi(\omega)\simeq 1+\xi'(1)(\omega -1)$. As we have
discussed the slope obtained
by this truncation will be overestimated, i. e.
the actual value should be more negative.

The RFT model must of course agree with
experimental data, but it also must be
consistent with fundamental results from QCD. The lattice simulation method
provides
several recent
results for $\xi'(1)$.
The calculation of
Bernard et al.
\cite{bib:bernard} gives
\be
\xi^{'}_{\bar{B}}(1)=-1.24 \pm 0.26\pm 0.33\ ,
\ee
while that of  Mandula and  Ogilvie \cite{bib:mandula}
yields
\be
\xi^{'}_{\bar{B}}(1)\simeq-0.95\ ,
\ee
and the UKQCD collaboration result \cite{bib:booth} is
\be
\xi^{'}_{\bar{B}}(1)=-1.2^{+0.7}_{-0.3}\ .
\ee
The agreement among these results is a gratifying demonstration of
the consistency of the RFT model with QCD.

\begin{center}
ACKNOWLEDGMENTS
\end{center}
This work was supported in part by the U.S. Department of Energy
under Contract No. DE-AC02-76ER00881 and in part by the University
of Wisconsin Research Commitee with funds granted by the Wisconsin Alumni
Research Foundation.

\begin{table}[p]
\begin{center}
\vspace{-1cm}
TABLES
\end{center}
\caption{ Heavy-light spin averaged states. Spin-averaged
masses are calculated in the usual way, by taking $\frac{3}{4}$ of
the triplet and $\frac{1}{4}$ of the singlet mass.}
\vskip 0.2cm
\begin{tabular}{|lccccc|}
\hline
\hline
state &\multicolumn{2}{c}{spectroscopic label\hspace{+2mm}\ \ \ } &
\ \ spin-averaged\ \ \  &\ \ \ theory\ \ \  &
\ \ \ error\ \ \ \\
      &      $J^{P}$        &  $^{2S+1}L_{J}$   &   mass (MeV)  & (MeV)    &
(MeV) \\
\hline
\underline{$c\bar{u},\ c\bar{d}$ quarks} & &   &               &           &
\\
$\begin{array}{ll}
   D & (1867) \\
   D^{*}&  (2010) \end{array}$ &
$\begin{array}{l}
      0^{-} \\
      1^{-} \end{array}$    &
$\left. \begin{array}{l}
\hspace{+1.1mm}   ^{1}S_{0} \\
\hspace{+1.1mm}   ^{3}S_{1} \end{array}\right] $  & $1S\ (1974)$ & $1975$ &
 $1$ \\
$\begin{array}{ll}
D_{1} & (2424) \end{array}$ & $
1^{+} $ & $\left.\begin{array}{l}
\hspace{+0.5mm}^{1}P_{1}\end{array}\right] $ &
$1P\ (2424)$ & $2425$ & $1$ \\
\underline{$c\bar{s}$ quarks} & & & & & \\
$\begin{array}{ll}
   D_{s} & (1969) \\
   D^{*}_{s}&  (2110) \end{array}$ &
$\begin{array}{l}
      0^{-} \\
      1^{-} \end{array}$    &
$\left. \begin{array}{l}
\hspace{+1.1mm}    ^{1}S_{0} \\
\hspace{+1.1mm}    ^{3}S_{1} \end{array}\right] $  & $1S\ (2075)$ & $2074$ &
 $-1$ \\
$\begin{array}{ll}
D_{s_{1}} & (2537) \end{array}$ & $
1^{+} $ & $\left.\begin{array}{l}
\hspace{+0.5mm}^{1}P_{1}\end{array}\right] $ &
$1P\ (2537)$ & $2536$ & $-1$ \\
\underline{$b\bar{u},\ b\bar{d}$ quarks} & & & & & \\
$\begin{array}{ll}
   B & (5279) \\
   B^{*} &  (5325) \end{array}$ &
$\begin{array}{l}
      0^{-} \\
      1^{-} \end{array}$    &
$\left. \begin{array}{l}
\hspace{+1.1mm}   ^{1}S_{0} \\
 \hspace{+1.1mm}   ^{3}S_{1} \end{array}\right] $  & $1S\ (5312)$ & $5311$ &
 $-1$ \\
\underline{$b\bar{s}$ quarks} & & & & & \\
$\begin{array}{ll}
B_{s} & (5368) \end{array}$ & $
0^{-} $ &
 $\left.\begin{array}{l} \hspace{+1.0mm} ^{1}S_{0}\end{array}\right] $ &
$1S\ (5409)$ & $5410$ & $1$ \\
\hline
\hline
\end{tabular}
\\
\begin{tabular}{ccclc}
\hspace{+5cm}$m_{u,d}$ &=& $300$& $MeV$ & (fixed) \\
\hspace{+5cm}$m_{s}$ &=& $513$& $MeV$&   \\
\hspace{+5cm}$m_{c}$ &=& $1292$&$ MeV$&  \\
\hspace{+5cm}$m_{b}$ &=& $4628$&$ MeV$&  \\
\hspace{+5cm}$a$ &=& $0.2$&$ GeV^{2}$&  (fixed) \\
\hspace{+5cm}$\kappa$ &=& $0.516$ & &
\end{tabular}
\label{tab:hl}
\end{table}

\begin{figure}[p]
\begin{center}
FIGURES
\end{center}
\vskip 0.01cm
\caption{IW function for $\bar{B}\ra D^{(*)}l\bar{\nu}_l$ decays
for $m_{u,d}=0$  and $m_{u,d}=0.3GeV$ (dashed line).}
\label{fig:iwfb}
\end{figure}

\begin{figure}[h]
\caption{Slope of the IW function at zero recoil for
$\bar{B}$ decays as a function of $m_{u,d}$.\hspace{+5cm}}
\label{fig:sl}
\end{figure}

\begin{figure}[h]
\caption{Second derivative of the IW function at zero recoil for
$\bar{B}$ decays as a function of $m_{u,d}$.\hspace{+5cm}}
\label{fig:sd}
\end{figure}

\begin{figure}[h]
\caption{Power series expansion about
$\omega =1$ (zero recoil point)
truncated at various
number of
terms. The solid curve is the full IW function
for $\bar{B}$ decays. In this figure
we have taken  $m_{u,d}=0.3\ MeV$.}
\label{fig:ps}
\end{figure}

\begin{figure}[h]
\caption{Ratio of $\xi_{\bar{B}_{s}}$ and $\xi_{\bar{B}}$ for
the
particular choice of $m_{u,d}=300MeV$. The mass of
the strange quark was $m_{s}=513MeV$ as a result of the
fit of Table \protect \ref{tab:hl}. We observe the two IW functions
differ by less than two percent
over the kinematically allowed range of $\omega$.}
\label{fig:iwfbs}
\end{figure}

\begin{figure}[h]
\caption{Comparison of the IW function
$\xi_{\bar{B}}(\omega)$ calculated from the RFT model
for the particular choice of $m_{u,d}=300\ MeV$
with  ARGUS \protect \cite{bib:arg}  and CLEO II
data \protect \cite{bib:cleo} for the decay
$\bar{B}^{0}\ra D^{*+}l^{-}\bar{\nu}_{l}$.
 For the sake of clarity, error bars
are shown only for the CLEO II data.}
\label{fig:arg}
\end{figure}

\begin{figure}[h]
\vskip 2cm
\end{figure}

\begin{figure}[p]
\epsfxsize = 5.4in \epsfbox{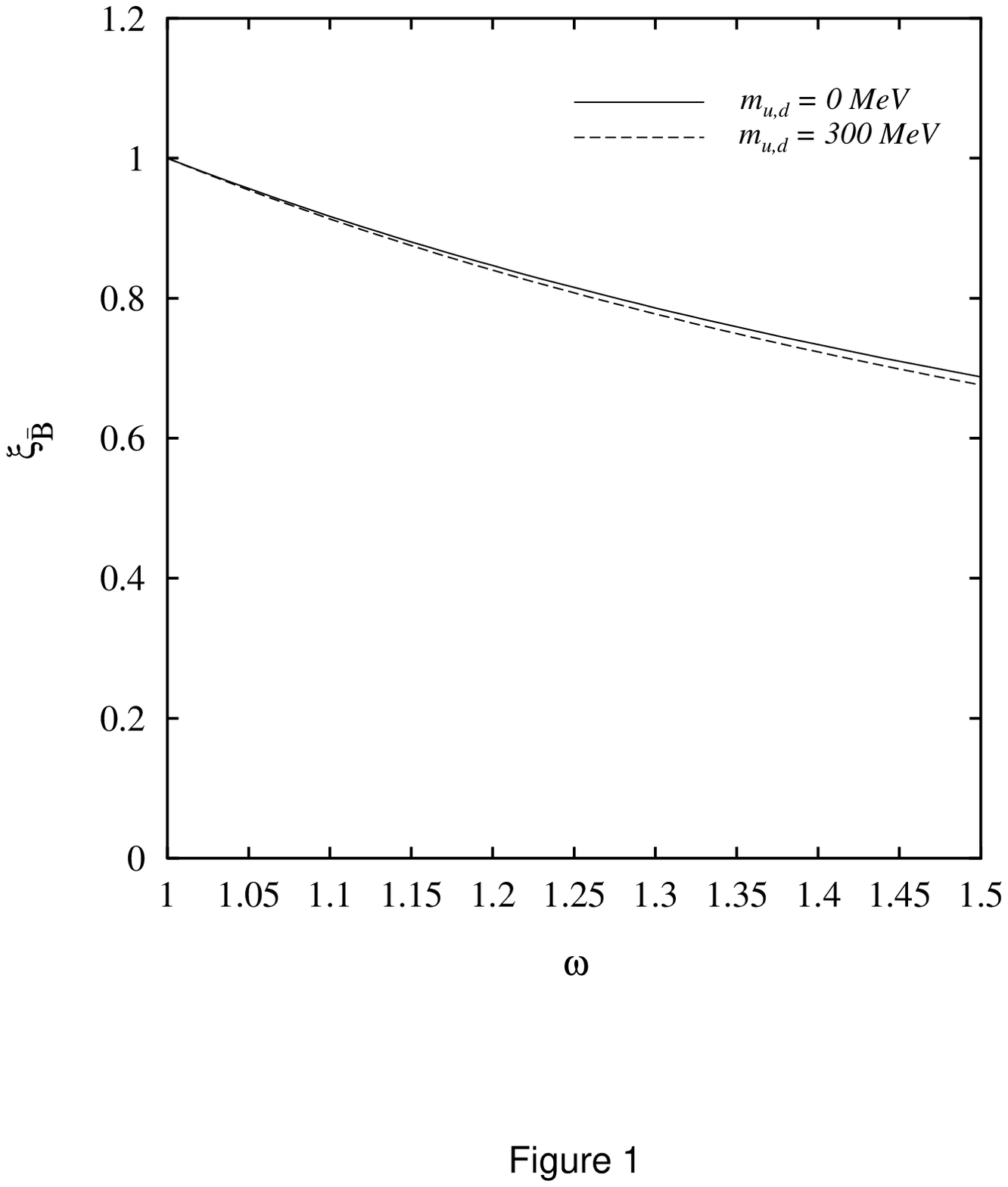}
\end{figure}

\begin{figure}[p]
\epsfxsize = 5.4in \epsfbox{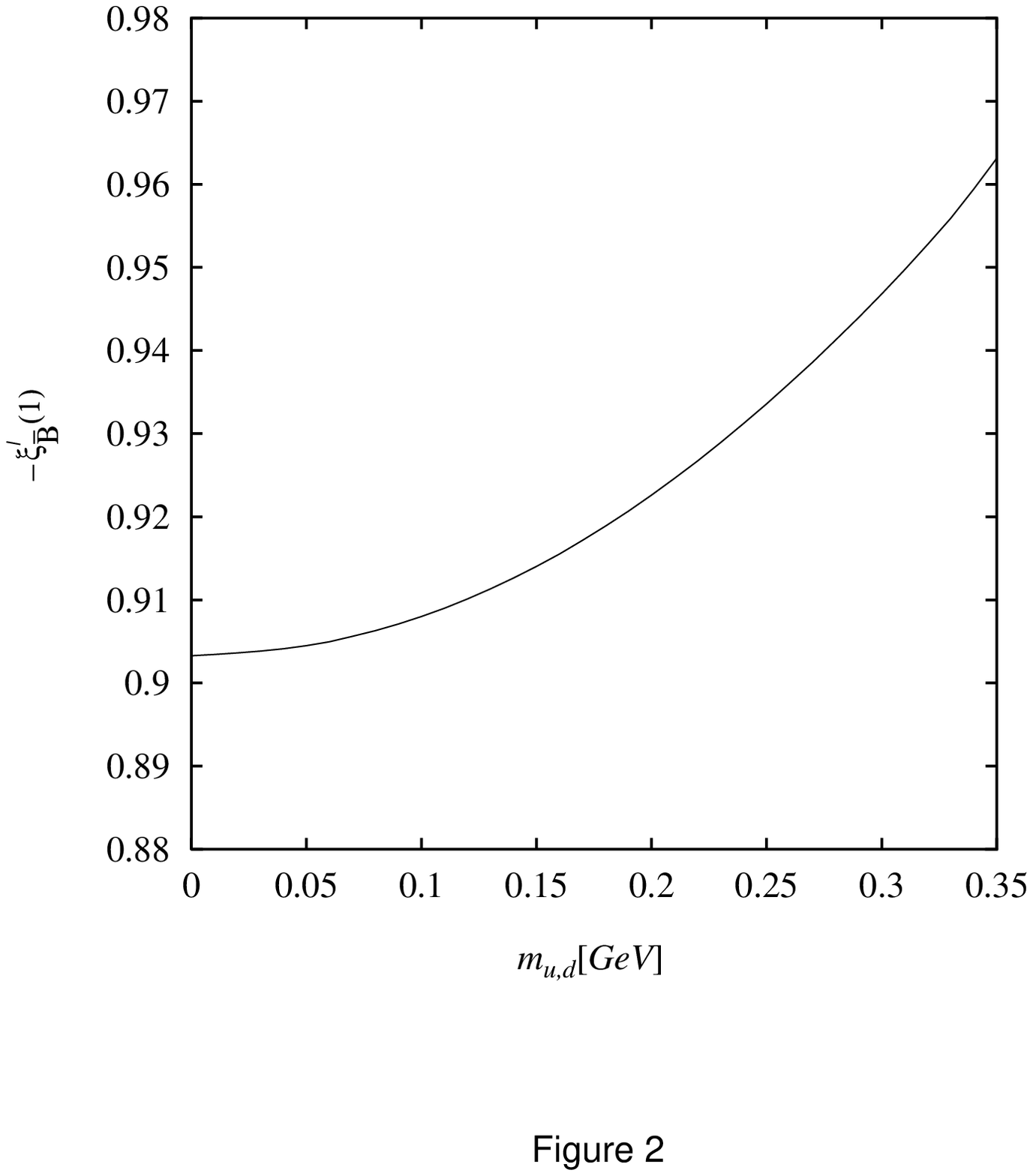}
\end{figure}

\begin{figure}[p]
\epsfxsize = 5.4in \epsfbox{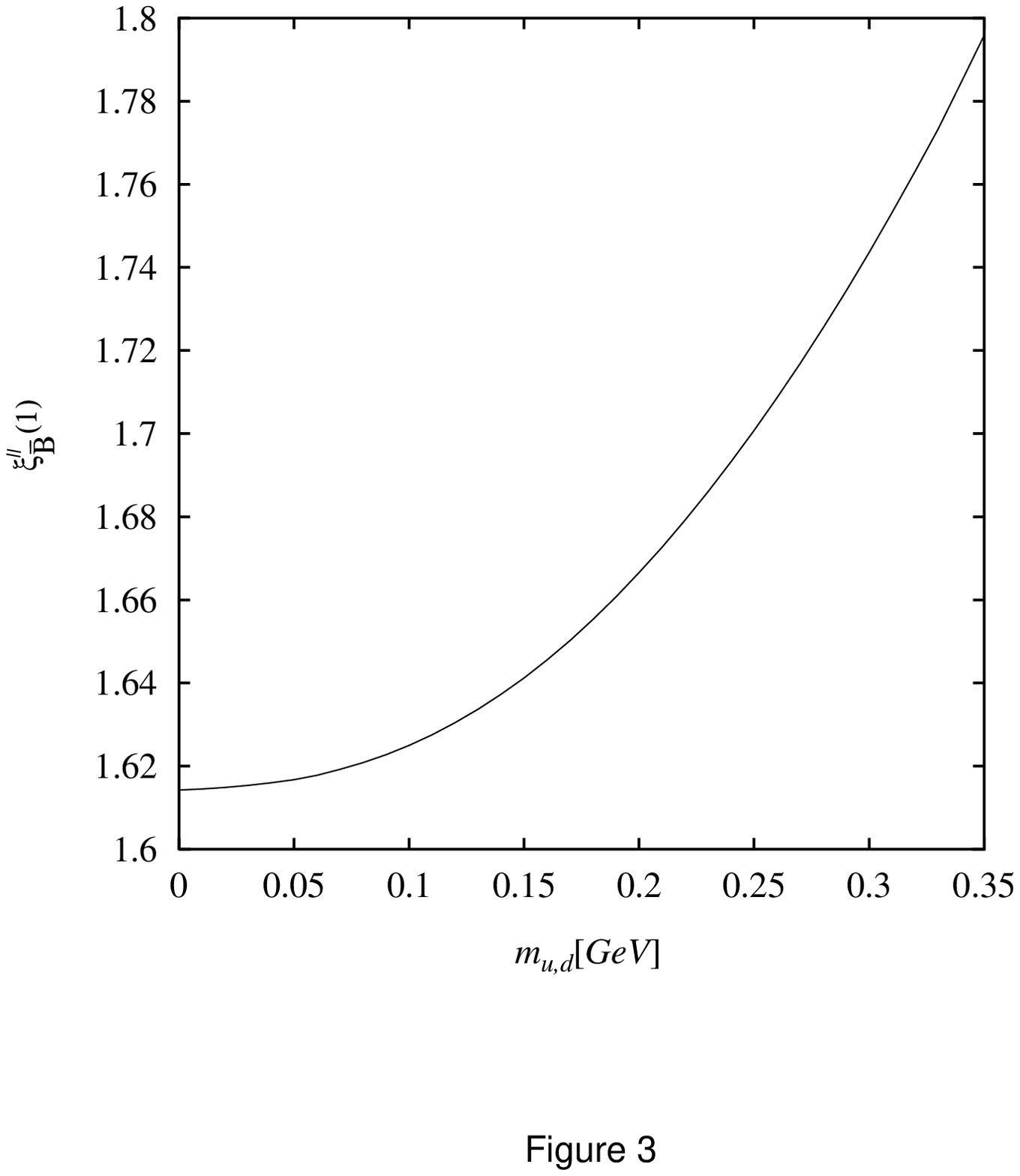}
\end{figure}

\begin{figure}[p]
\epsfxsize = 5.4in \epsfbox{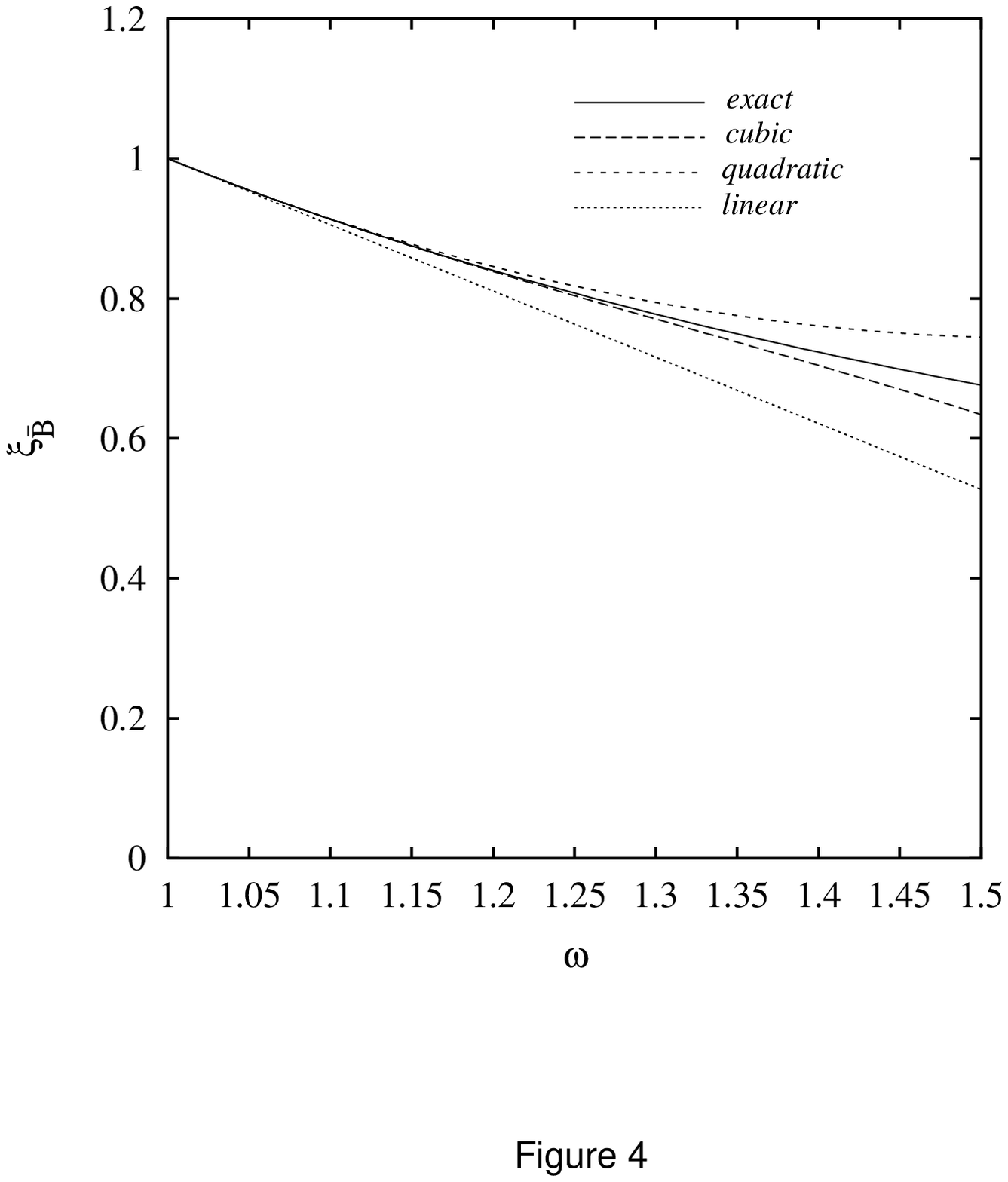}
\end{figure}

\begin{figure}[p]
\epsfxsize = 5.4in \epsfbox{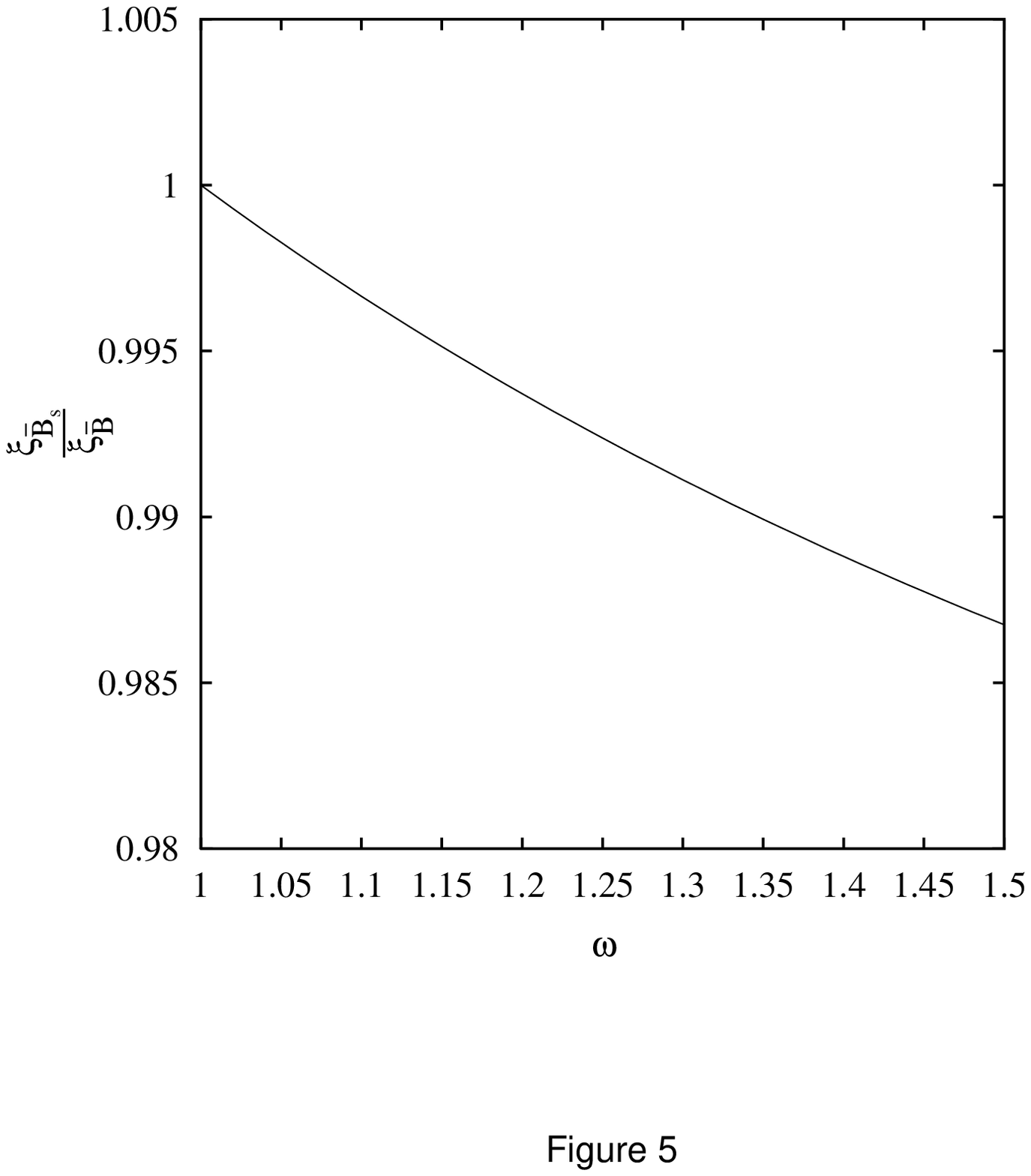}
\end{figure}

\begin{figure}[p]
\epsfxsize = 5.4in \epsfbox{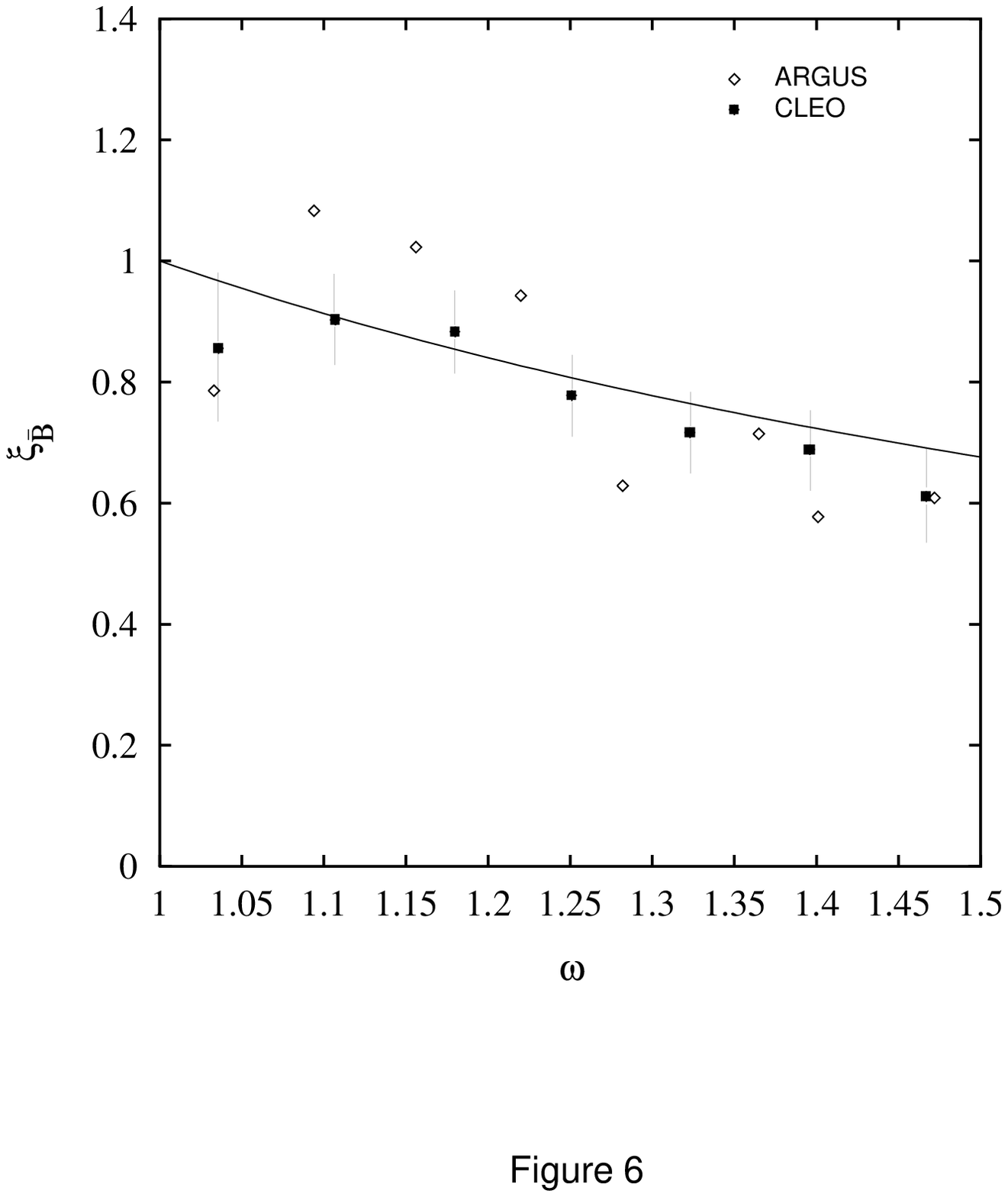}
\end{figure}


\newpage
\begin{thebibliography}{99}
\bibitem{bib:isg-wis} N. Isgur and M. B. Wise,
Phys. Lett. {\bf 232B}, 113 (1989); {\bf 237B}, 527 (1990).
\bibitem{bib:dan} Dan LaCourse and M. G. Olsson,
Phys. Rev. D {\bf 39}, 2751 (1989).
\bibitem{bib:col-ken} Collin Olson, M. G. Olsson and Ken Williams,
Phys. Rev. D {\bf 45}, 4307 (1992).
\bibitem{bib:ken} M. G. Olsson and Ken Williams,
Phys. Rev. D {\bf 48}, 417 (1993).
\bibitem{bib:col-dan} C. Olson, M. G. Olsson and D. LaCourse,
Phys. Rev. D. {\bf 49}, 4675 (1994).
\bibitem{bib:sik} M. G. Olsson and Sini\v{s}a Veseli,
{\em The Asymmetric Flux Tube}, UW-Madison preprint MAD/PH/816.
\bibitem{bib:ei-fe} E. Eichten and F. Feinberg,
Phys. Rev. D {\bf 23}, 2724 (1981);
D. Gromes,
Z. Phys. C {\bf 22}, 265 (1984); {\bf 26}, 401 (1984);
A. Barchielli, E. Montaldi and G. M. Prosperi,
Nucl. Phys.  {\bf B296}, 625 (1988); {\bf B303}, 752(E) (1988).
\bibitem{bib:ba-br-pr} A. Barchielli, N. Brambilla and G. M. Prosperi,
Nuovo Cim  {\bf 103A}, 59 (1989).
\bibitem{bib:br-pr} N. Brambilla and G. M. Prosperi,
Phys. Lett.  {\bf 236B}, 69 (1990).
\bibitem{bib:buch} W. Buchm$\ddot{\rm u}$ller,
Phys. Lett. {\bf 112B}, 479 (1982).
\bibitem{bib:pi-st} Robert D. Pisarski and John D. Stack,
Nucl. Phys.  {\bf 286B}, 657 (1987).
\bibitem{bib:galerkin} Steve Jacobs, M. G. Olsson, and Casimir Suchyta III,
Phys. Rev. D {\bf 33}, 3338 (1986).
\bibitem{bib:li-ho} O. Lie-Svendsen
and H. H\hspace{-1.25mm}$\not\hspace{-0.35mm}{o}$gaasen,
Z. Phys. C {\bf 35}, 239 (1987).
\bibitem{bib:sa-za} M. Sadzikowski and K. Zalewski,
Z. Phys. C {\bf 59}, 677 (1993).
\bibitem{bib:aglietti} U. Aglietti, G. Martinelli, and C. T. Sachrajda,
Phys. Lett. {\bf B324}, 85 (1994).
\bibitem{bib:arg} H. Albrecht et al., ARGUS Collaboration,
Z. Phys. C {\bf 57}, 533 (1993).
\bibitem{bib:cleo} B. Barish et al., CLEO Collaboration,
{\em Measurement of the $\bar{B}\ra D^{*}l\bar{\nu}_{l}$
anti-neutrino branching ratios and $|V_{cb}|$},
Cornell Nuclear Studies Wilson Lab preprint, HEPEX 9406005 (1994).
\bibitem{bib:neudif} M. Neubert,
Phys. Lett. {\bf 264B}, 455 (1991).
\bibitem{bib:pdg} Particle Data Group,
Phys. Rev. D {\bf 50}, Part I (1994).
\bibitem{bib:slopes} G. V. Efimov et al.,
Z. Phys. C {\bf 54}, 349 (1992);
A. V. Radyushkin,
Phys. Lett {\bf 271B}, 218 (1991).
\bibitem{bib:bernard} C. Bernard, Y. Shen and A. Soni,
Phys. Lett {\bf 317B}, 164 (1993).
\bibitem{bib:mandula} J. E. Mandula and M. C. Ogilvie
{\em A Lattice Calculation of the Heavy Quark Universal
Form Factors}, HEPLAT 9312013 (1993).
\bibitem{bib:booth} S. P. Booth at al.,
Phys. Rev. Lett. {\bf 72}, 462 (1994).
\end{thebibliography}
\end{document}